\documentclass[%
 reprint,
superscriptaddress,
 amsmath,amssymb,
 aps,
pre,
]{revtex4-2}

\usepackage{epsfig,dsfont,amssymb,amsmath,amsthm,amsfonts,amsbsy,mathrsfs}
\usepackage{graphicx}
\usepackage{color}
\usepackage{bm}
\usepackage{multirow}
\usepackage{natbib}
\usepackage{comment}

\usepackage{tikz}
\usetikzlibrary{arrows,shapes,chains}

\newcommand{\rr}{{\mathbf r}}

\newcommand{\BEQ}{\begin{equation}}
\newcommand{\EEQ}{\end{equation}}
\newcommand{\BEA}{\begin{eqnarray}}
\newcommand{\EEA}{\end{eqnarray}}

\newcommand{\ee}{\mathbf{e}}

\begin{document}
\preprint{APS/123-QED}


\title{Epidemic processes on self-propelled particles: continuum and agent-based modelling}



\author{Jorge P. Rodríguez}
\affiliation{Instituto Mediterráneo de Estudios Avanzados IMEDEA (CSIC-UIB), C/ Miquel Marqu\`es 21, 07190 Esporles, Spain}
\affiliation{Institute for Biocomputation and Physics of Complex Systems (BIFI), University of Zaragoza, 50018 Zaragoza, Spain}
\author{Matteo Paoluzzi}
\email{matteopaoluzzi@ub.edu}
\affiliation{Departament de Física de la Mat\`eria Condensada, Universitat de Barcelona, Martí Franqu\`es 1, 08028 Barcelona, Spain \\}
\author{Demian Levis}
\affiliation{Departament de Física de la Mat\`eria Condensada, Universitat de Barcelona, Martí i Franqu\`es 1, 08028 Barcelona, Spain \\}
\affiliation{ Universitat de Barcelona Institute of Complex Systems (UBICS), Martí Franqu\`es 1, 08028 Barcelona, Spain \\}
\author{Michele Starnini}
\email{michele.starnini@gmail.com}
\affiliation{ISI Foundation, via Chisola 5, 10126 Torino, Italy}
\affiliation{Departament de F\'isica i Enginyeria Nuclear, Universitat Polit\'ecnica de Catalunya, Campus Nord B4, 08034 Barcelona, Spain}


\begin{abstract}
Most spreading processes require spatial proximity between agents. The stationary state of spreading dynamics in a population of mobile agents thus depends on the interplay between the time and length scales involved in the epidemic process and their motion in space. We analyze the steady properties resulting from such interplay in a simple model describing epidemic spreading (modelled as a Susceptible-Infected-Susceptible process) on self-propelled particles (performing Run-and-Tumble motion). 
Focusing our attention on the diffusive long-time regime, we find that the agents' motion changes qualitatively the nature of the epidemic transition  characterized by the emergence of a macroscopic fraction of infected agents. Indeed, the transition becomes of the mean-field type for agents diffusing in one, two and three dimensions,  while, in the absence of motion, the epidemic outbreak depends on the dimension of the underlying static network determined by the agents' fixed locations. 
The insights obtained from a continuum description of the system are validated by numerical simulations of an agent-based model.
Our work aims at bridging  soft active matter physics and theoretical epidemiology, and may be of interest for researchers in both communities.
%
%
\end{abstract}

\maketitle

\section{Introduction}


The study of spreading processes on mobile agents is a field attracting growing interest in both communities of epidemiology and active matter physics.
On the one hand, human mobility plays a crucial role in the spreading of infectious diseases, as shown by the inclusion of mobility data into epidemic forecasting \cite{Balcan2009}.
Short range mobility -- such as individuals walking in a limited space -- has also been taken into account for epidemic modelling \cite{GONZALEZ2004741}, in particular by considering a Susceptible-Infected-Recovered (SIR) model in a population of random walkers \cite{frasca2006dynamical, buscarino2008disease}.
Furthermore, the interplay between mobility and spreading dynamics can be used to model behavior change in individuals \cite{ventura2022epidemic}, as well as to show that a feedback mechanism between the epidemic status and the agent's motion can enhance the contagion dynamics, effectively reducing the epidemic threshold \cite{PhysRevResearch.2.032056}.

On the other hand, classical spreading process can model well the diffusion of information in a population:  individuals aware of the information (infected) transmit it to unaware (susceptible) peers \cite{castellano2009statistical}.
Such information exchange is mediated locally by social interactions, involving agents with physical proximity, as observed also in the animal reign \cite{lusseau2009emergence,doi:10.1098/rsbl.2004.0225, doi:10.1098/rspb.2004.3019,croft_book, Croft2005}.
As a consequence, populations of motile, self-propelled agents self-organise in time and space, with the emergence of coordination and collective behavioral change \cite{Rosenthal4690}. 
Examples range from multi-cellular organisms to flocks of birds \cite{SumpterBook}, robot swarms \cite{Rubenstein2014}, or the coherent motion of fish schools avoiding a predator's attack \cite{ioannou2012predatory}. 
Also bacteria, which communicate through chemical signals that regulate their motion, show coordinated behavior of the whole population  \cite{keller2006communication, hibbing2010bacterial}, a mechanism know as quorum sensing \cite{waters2005quorum}.

From a physics standpoint, systems of self-propelled agents are typically modelled as persistent random walkers \cite{berg2018random, romanczuk2012active,solon2015active}. A salient example is the Run-and-Tumble (RnT) walk, which mimics the motion performed by several flagellated bacteria species such as Escherichia Coli \cite{berg2004coli}. Collectives of such self-propelled particles have been the focus of intense research in the past decades, providing a natural playground to explore living matter from a physics perspective \cite{Marchetti13, roadmap, needleman2017active}. These so-called Active systems, made of biomimetic entities, exhibit a remarkable richness of non-equilibrium collective states as a result of different kinds of interactions \cite{smeets2016emergent,PhysRevResearch.1.023026,PhysRevX.7.011028,PhysRevE.97.042604}, which can be of very different nature, say, mechanical, 'social', chemical, etc. Understanding  how  self-propulsion  changes  the  qualitative features of spreading dynamics and how a local information spreads in a collection of moving entities remains a fundamental open problem in the field.  
A feedback loop that couples motility with local density can be employed in experiments for controlling colloids \cite{bauerle2018} or photokinetic bacteria \cite{frangipane18,poon}, and it has been found that  Susceptible-Infected-Susceptible (SIS) dynamics can be employed for driving pattern formation and collective motion in systems of active particles  \cite{PhysRevResearch.2.032056, paoluzzi2020information,PhysRevE.98.052603}.

Therefore, it is of interest to shed light on the emergent behavior resulting from the interplay between motion and spreading dynamics.
Within this framework, some works have considered modelling the exchange of information in populations of mobile agents by the introduction of an extra internal degree of freedom controlled by an epidemic local process \cite{peruani2008dynamics, gonzalez2006system,PeruaniLee,rodriguez2019particle, peruani2019reaction, norambuena2020understanding, zhao2021contagion, PhysRevResearch.2.032056, ventura2022epidemic}.
Here, we introduce a simple model aiming at studying  spreading in systems of motile agents at a fundamental level, under both agent-based and continuum frameworks.
In particular, our approach allow us to show that motion, here in the form of enhanced diffusion, generically leads to a  homogeneous spreading across the population, independently of the space dimension where agents move. As agents diffuse faster, the epidemic threshold is reduced yet the nature of the transition is generically of the mean-field type.

\section{Susceptible-Infected-Susceptible dynamics on active particles}

We consider a system of $N$ active agents, or particles, that can take  two different internal states, labeled as Susceptible (S) and Infected (I). 
Each of them performs an independent RnT motion \cite{Schnitzer}:  a sequence of ‘‘runs’’ -- straight-line motion at  speed $v$ -- interrupted by ‘‘tumbles’’ -- random re-orientations of the self-propulsion direction -- occurring at a rate $\alpha$. Here the tumbling rate $\alpha_{S,I}$ and self-propulsion velocity $v_{S,I}$,  might depend   on the internal state of the agent, being  S or I.

\begin{figure}
\centering
\includegraphics[width=0.47\textwidth]{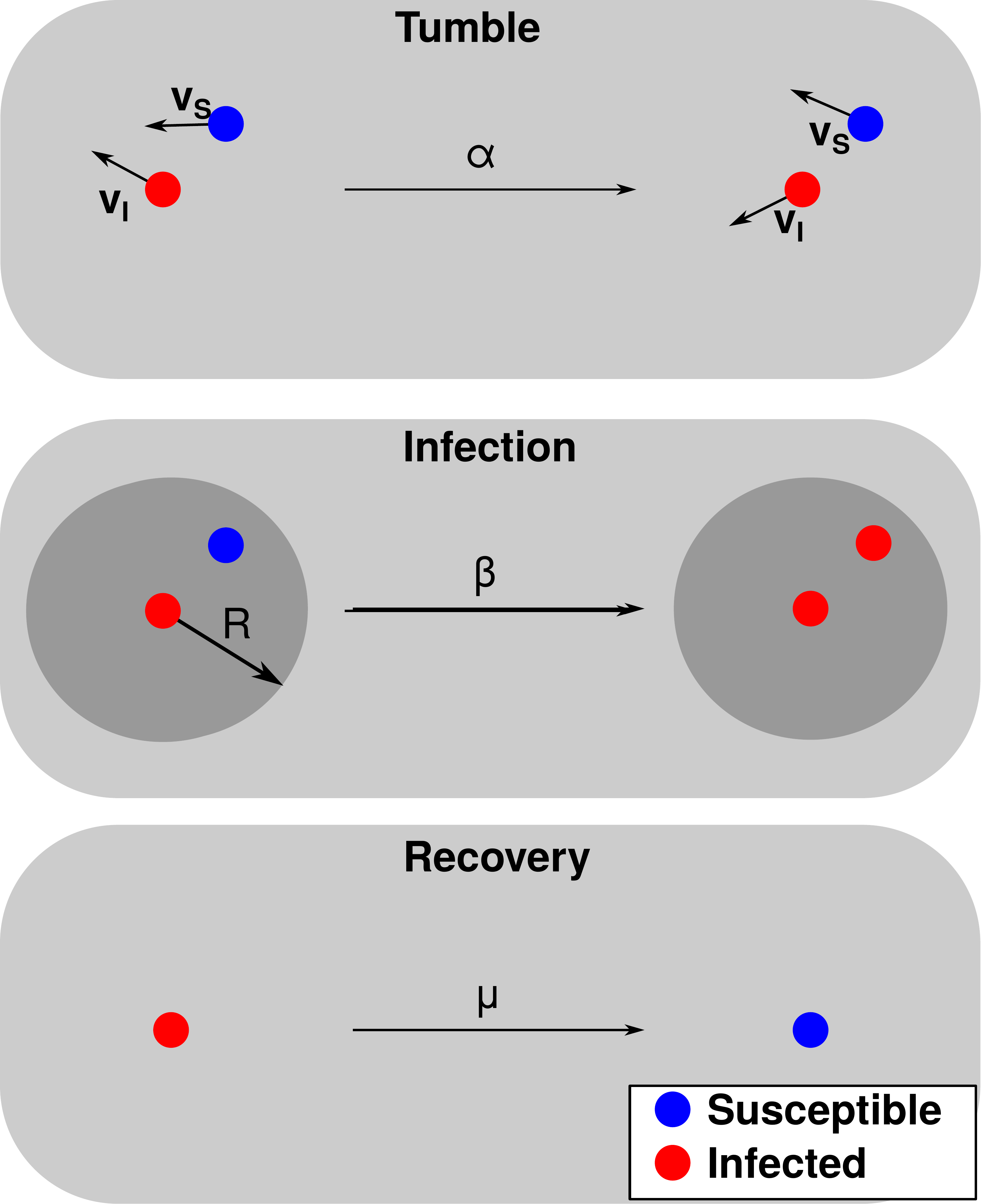}
\caption{Illustration of the dynamics of the model for $d=2$, made of three different stochastic processes: the tumbles, the infection and the recovery. 
Top panel: Susceptible (Infected) particles, depicted in blue (red), perform  RnT motion with tumbling rate $\alpha_S$ ($\alpha_I$) and self-propulsion velocity $v_S$ ($v_I$).
Middle panel: A S particle, in blue, located within the interaction radius $R$ of an I particle, in red, gets infected with rate $\beta$.
Bottom panel: I particles spontaneously recover (become S) with rate $\mu$.
}
\label{figure0}
\end{figure}

We  consider a $d$-dimensional  system where $N$ particles move in a 
 $L^d$ box with periodic boundary conditions.
The position of the particles $\rr_i(t)$ evolves according to 
\begin{align}
    \dot{\rr}_i^{S,I} (t)= v_{S,I} \ee_i (t)
\end{align}
with $\ee_i (t)$ indicating the unit vector that specifies the swimming direction at time $t$, which changes randomly at a rate $\alpha_{I,S}$ (see top panel of Figure \ref{figure0}).

The agents' internal states are then subjected to a SIS process \cite{siroriginal}, where the transition rates between states are defined as usual 
\begin{equation}
 S+I \xrightarrow{\beta} 2I \,, \, \ I \xrightarrow{\mu} S.
  \end{equation}
The $S+I \xrightarrow{\beta} 2I$ reaction takes place with rate $\beta$ only in regions of space where particles $S$ and $I$ overlap, that is, exclusively for those pairs of particles located
within a distance $R$, chosen to be small with respect to the size of the system $R \ll L$ (see middle panel of Figure \ref{figure0}).
Conversely, infected particles decay spontaneously to the susceptible state with rate $\mu$ (see bottom panel of Figure \ref{figure0}).

Here, we focus on the case $\beta>0$ and $\mu>0$. 
For $\mu=0$ and $\beta>0$, the system evolves towards an absorbing state were  all particles are infected in the so-called SI dynamics. It has been shown that, even in this case -- trivial from the point of view of spreading dynamics -- the interplay between the collision rate and the persistence length triggers fractal growth in the diffusive limit \cite{PhysRevE.98.052603}. 
Non-zero values of $\mu$ {and large $\beta$ values} trigger a mixed phase, where active particles can develop patterns, while for large values of $\mu$ (when the recovery rate is faster than the collision rate), the system can eventually evolve into an absorbing configuration of susceptible particles \cite{paoluzzi2020information}.

In the following, we  will analyze the system under both a continuum and agent-based perspective. For the latter, we will perform numerical simulations. 
At each time step $t$, the interactions between particles are described by a spatial network where nodes represent particles and links represent interactions between them (thus, two particles are linked if they are within a distance $R$). 
As particles move, the set of links  evolves.
This temporal network can  thus be described by a sequence of snapshots, each one with  mean connectivity $\langle k \rangle$ \cite{temporalnetworksbook}.


We run numerical simulations by implementing the Gillespie algorithm \cite{gillespie}. 
The time required to update the system is in this case a stochastic variable to be sampled. Once the time needed to perform a given update has been computed, the system is updated. The time associated to a given update  is generated from an exponential distribution with characteristic time given by the inverse of the sum of all the transition rates involved in the evolution of our system.  Then, one of these transitions, or processes (reactions in the spirit of Gillespie), is chosen with a probability proportional to its rate. Such processes are: 

\begin{itemize} 

\item Run: We update the particles' positions at a rate $\lambda_m=5000$, much higher than any of the other dynamic processes in our model, such that the particle movement can be considered continuous (to be compared with the rates of the other processes involved : $\alpha$, $\beta$, $\mu \lesssim 1$). The positions of all particles are updated synchronously.
\item Tumble: Particles change their  direction of motion at a rate $\alpha$. All the particles' directions are updated synchronously. 
\item Infection: 
The rate of infection of a susceptible particle $i$ at time $t$ is $\beta I_i(t)$, where $I_i(t)$ is the number of neighbouring infected particles  at time $t$.
\item Recovery: Infected nodes recover with rate $\mu$.
\end{itemize}
Therefore, the total updating rate of the system is $\delta(t)=\lambda_m+\alpha+\beta\sum_{i \in \text{S}}I_i(t)$+$\mu N \rho(t)$, where the sum runs over susceptible particles and $\rho(t)$ is the prevalence, that is, the fraction of infected particles in the system, at time $t$. 

We fix $N=1024$, and the time and length unit by setting $\mu=1$ and $L=1$. 
We consider that, for each value of $\beta$, 1 Monte Carlo (MC) step corresponds to the time it would take to reach the state where all agents are infected under exponential growth $\rho(t)=e^{\beta \langle k \rangle t}$, leading to $1\text{ MC}=\frac{\log N}{\beta \langle k \rangle}$. We start our simulations from a disordered distribution of agents at high values of $\beta$. We then let the system relax 60 MC steps and use the final steady configuration as the initial configuration for a new simulation at a lower value of $\beta$. We let the system relax 10 MC steps, and then we repeat this procedure by subsequently reducing the value of $\beta$ to eventually reach the steady configurations at each value of $\beta$. 

As a reference, we will consider two limit regimes based on a time-scale separation between motion and spreading: (i) the static regime where particles do not move (or move at much longer time scales than the spreading process), and (ii) the homogeneous-mixing regime, where particles move at much shorter time scales than the ones involved in the spreading. In both limits, the only two control parameters are  the infection and recovery rates, as the positions are either not updated or updated at random. In the homogeneous-mixing limit the exact steady density of infected particles (or prevalence) is $\rho = 1-\frac{1}{\beta \langle k \rangle}$ for $\beta \langle k \rangle > 1$, and $\rho=0$ otherwise, resulting in an epidemic threshold $\beta_c\langle k\rangle=1$ beyond which $\rho>0$.

\section{Continuum model}

In order to gain insight into the large-scale behavior of the system described so far, we adopt a continuum approach by considering the SIS dynamics on top of the run-and-tumble master equation \cite{Schnitzer}.  
We introduce $\tilde{S}(\rr,\ee,t)$ and $\tilde{I}(\rr,\ee,t)$ as the probability density function of, respectively, susceptible and infected particles at location $\rr$ with orientation $\ee$ at the time $t$. We are interested in the time-evolution of $S(\rr,t)=\int \text{d}\ee \, \tilde{S}(\rr,\ee,t)$ and $I(\rr,t)=\int \text{d}\ee \, \tilde{I}(\rr,\ee,t)$. 
We can associate to $S(\rr,t)$ and $I(\rr,t)$ the currents $\mathcal{J}_{I,S}(\rr,t)$ that are defined as $\mathcal{J}_{S}(\rr,t) = v_S\int \text{d}\ee \,\ee \tilde{S}(\rr,\ee,t)$ and $\mathcal{J}_{I}(\rr,t) = v_I\int \text{d}\ee \,\ee \tilde{I}(\rr,\ee,t)$. 
Focusing our attention on the case where the motility parameters $\alpha_{I,S}$ and $v_{I,S}$ might dependent on the internal state of the agents but homogeneous in space, we obtain that
the dynamics of the concentration fields $\tilde{S}$ and $\tilde{I}$ is governed by the following equations
\begin{align}
    \partial_t \tilde{S} &= -v_S \nabla \cdot (\boldsymbol{e} \tilde{S}) - \alpha_S \mathcal{Q} [\tilde{S}] - \beta I \tilde{S} + \mu I \\ 
        \partial_t \tilde{I} &= -v_I \nabla \cdot (\boldsymbol{e} \tilde{I}) - \alpha_I \mathcal{Q} [\tilde{I}] + \beta S \tilde{I} - \mu I 
\end{align}
where we have introduced the projector operators $\mathcal{Q}\equiv 1 - \mathcal{P}$. The operator $\mathcal{P}$ acts on a generic function, integrating over all possible directions $\boldsymbol{e}$ of motion \cite{Schnitzer}. We denote it as follows
\begin{align}
    \mathcal{P}[f] &\equiv \frac{1}{\Omega} \int \text{d}\boldsymbol{e} \, f \\ \nonumber 
    \Omega &\equiv \int \text{d}\boldsymbol{e} \; 
\end{align}
where $f\!=\!f(\boldsymbol{e},\boldsymbol{r},t)$.
Within this notation, the average prevalence (the concentration  of infected particles) in the system at time $t$ is given by $\rho(t) = \int I(\rr,t) \text{d} \rr$.

In this way, 
we can write the following set of equations for the densities and their currents
\begin{subequations}\label{eq_mod_0}
\begin{align}
    \dot{S} &= -\nabla \cdot \mathcal{J}_S + f(S,I) \\ 
    \dot{I} &= - \nabla \cdot \mathcal{J}_I - f(S,I) \\ 
    \dot{\mathcal{J}}_S &= -\frac{v^2_S}{d} \nabla S - \mathcal{J}_S \left[ \alpha_S + \beta I \right] \\
    \dot{\mathcal{J}}_I &= -\frac{v^2_I}{d} \nabla I - \mathcal{J}_I \left[ \alpha_I -\beta S\right] \\ 
        f(S,I) &\equiv -\beta I(S - \frac{\mu}{\beta}) \; .
\end{align}
\end{subequations}

We now focus our attention on equations for the currents. Without loss of generality,  let us discuss the equation for the current of $S$. As a first choice, we can consider that the active dynamics define the relevant time scale through the tumbling rate $\alpha_S$ and thus write
\begin{align}
    \alpha_S^{-1}\dot{\mathcal{J}}_S &= -\frac{v^2_S}{d \alpha_S} \nabla S - \mathcal{J}_S \left[ 1 + \frac{\beta}{\alpha_S} I \right]\,.
\end{align}
Active particles will reach stationarity on time scales $t \gg \alpha_S^{-1}$. We can thus consider safely a diffusive limit obtained by considering $\alpha_S^{-1} \to 0$, $v_S \to \infty$ in a way that we are keeping  $D_S^0 \equiv v_S^2/d \alpha_S$ fixed  \cite{kac1974stochastic}. Following the same trend of ideas for the current of $I$, in this limit of vanishing currents, i. e., $\dot{\mathcal{J}}_{S,I}=0$, we get the following constitutive relations
\begin{align}
    \mathcal{J}_{S} = -D^0_{S} \nabla S \\ 
    \mathcal{J}_{I} = -D^0_{I} \nabla I \; .
\end{align}
In  this  limit, we are assuming active particles reach a stationary state before the spreading process. In this picture, we obtain that the dynamics of the system is captured by a two-component reaction-diffusion process conserving the total mass $\overline{\rho}(t) =\int d\rr \left[ S(\rr,t) + I(\rr,t) \right]$  \cite{PhysRevX.10.041036,halatek2018rethinking} and is described by the following equations 
\begin{align} \label{eq:model1}
\partial_t S(\rr,t) &= D^0_S \nabla^2 S + f(S,I) \\ \label{eq:model2}
\partial_t I(\rr,t) &= D^0_I \nabla^2 I - f(S,I) \; ,
\end{align}
where suitable boundary conditions have to be taken into account. 

Another limiting case can be obtained considering that the spreading process is much faster than the RnT motion. In this limit, the spreading dynamics reaches a stationary state before active particles are able to reach the diffusive limit. For studying this limiting situations it turns out convenient to write the equations for the currents in the following way  
\begin{align}
        \beta^{-1} \dot{\mathcal{J}}_S &= -\frac{v^2_S}{\beta d} \nabla S - \mathcal{J}_S \left[ \frac{\alpha_S}{\beta} + I \right] \\
    \beta^{-1}\dot{\mathcal{J}}_I &= -\frac{v^2_I}{\beta d} \nabla I - \mathcal{J}_I \left[ \frac{\alpha_I}{\beta} - S\right] 
\end{align}
In this  case, once we define $\tilde{D}_{S,I}\equiv v_{S,I}^2/\beta d$, we obtain the following equations
\begin{align}
    \partial_t S &=  \tilde{D}_S \nabla (\frac{1}{I} \nabla S) + f(S,I) \\ 
    \partial_t I &= -\tilde{D}_I \nabla (\frac{1}{S} \nabla I) - f(S,I) \; .
\end{align}
As one can see, the equation for $I$ has the form of a backward diffusion equation that tends to make the $I$ profile less smooth as time increases. 
In the limit $\beta \to \infty$ or $v_{S,I} \to 0$, one has $\tilde{D}_{S,I}=0$ and thus the spreading dynamics involves only regions where the two density fields overlap.

Away from these two limiting situations, we can still look for stationary solutions that are obtained considering vanishing currents.
These solutions provide the constitutive relations $\mathcal{J}_{S,I}=\mathcal{J}_{S,I}[S,I]$ that once plugged into the equations for $S$ and $I$ bring to the following evolution equations %
%
\begin{subequations}\label{eq_mod}
\begin{align} 
    \partial_t S &= \nabla \left[\mathcal{D}_s(I) \nabla S \right]+ f(S,I) \\ 
    \partial_ t I &= \nabla \left[\mathcal{D}_I(S) \nabla I \right]- f(S,I) \\ 
    \mathcal{D}_S(I) & = D^0_S \frac{1}{1 + \frac{\beta I}{\alpha_S}} \\ 
    \mathcal{D}_I(S) & = D^0_I \frac{1}{1 - \frac{\beta S}{\alpha_I}} \; .
\end{align}
\end{subequations}
The functional form of this equations show that the interplay between RnT dynamics and spreading process makes the 
effective diffusion constant $\mathcal{D}_{S,I}$ space-and state-dependent. Meaning that, if we color in different way $S$ and $I$ agents, Eqs. (\ref{eq_mod}) suggest that, while agents' motion is diffusive with diffusivity $D^0_{S,I}$, fluctuations of color are spatially heterogeneous.

\section{From static to {homogeneous} mixing}

We now consider RnT motion with  $v_S\!=\!v_I\!=\!v$ and $\alpha_S\!=\!\alpha_I\!=\!\alpha$.
Thus, both states are described by the same diffusion constant $D_S^0 = D_I^0 =D $.
Next, we describe the SIS dynamics in the three main cases of active particles motion: i) the static limit in which particles do not move, $D=0$, ii) the homogeneous mixing limit in which particles move arbitrarily fast and are well described by mean-field, $D \rightarrow \infty$, and iii) the crossover between static and mean-field limits.

\begin{figure}
\includegraphics[width=0.47\textwidth]{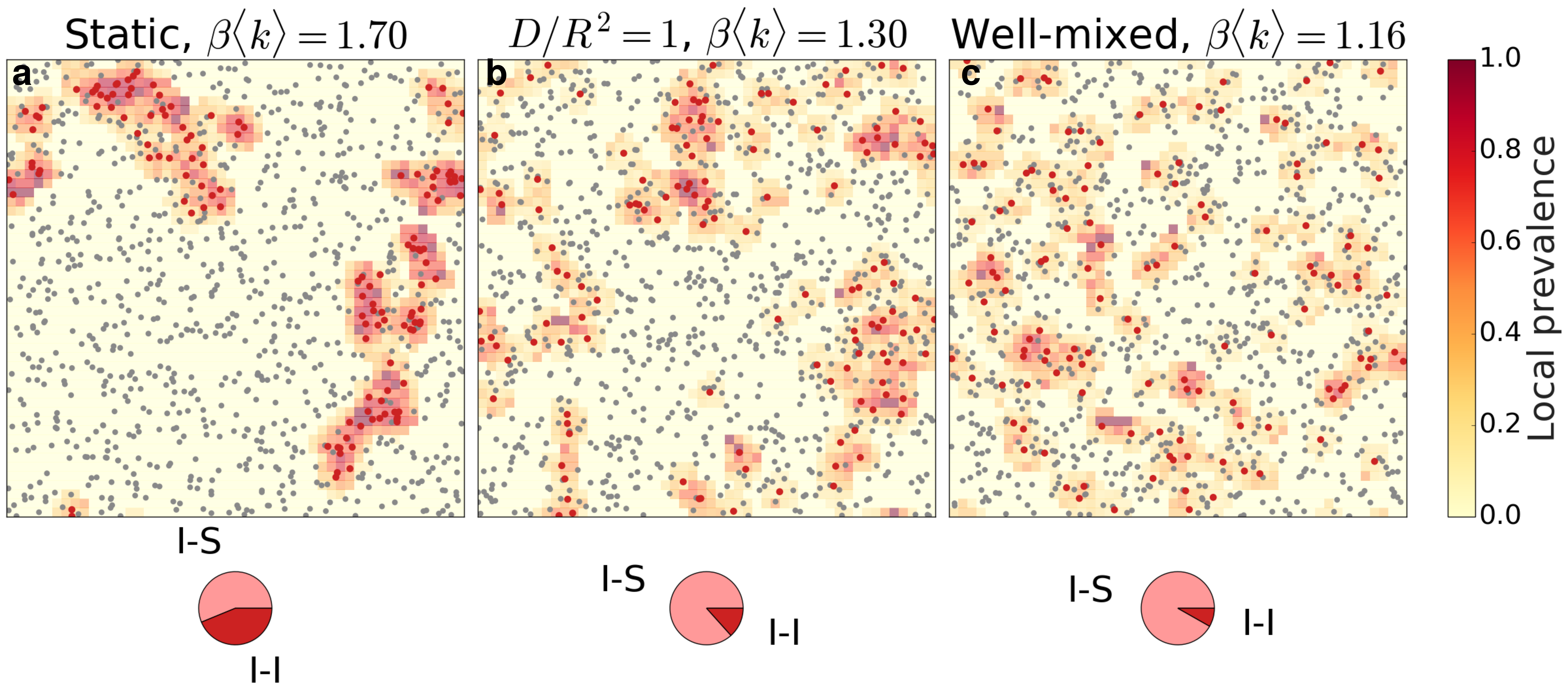}
\caption{ Snapshots of three steady configurations with the same prevalence $\rho=0.13$ but different diffusion constant $D$: a) static case $D=0$, b) $D/R^2=1$  and c) well-mixed case $D \rightarrow \infty$. Colors indicate the state of particles: susceptible (grey) and infected (red). The background color corresponds to the value of the local prevalence. 
Pie charts at the bottom illustrate the number  I-S (infected-susceptible) and I-I (infected-infected) links.}
\label{fig:confs}
\end{figure}

Figure \ref{fig:confs} shows three snapshots of stationary 
configurations corresponding to a) the static case ($D = 0$), b) diffusive limit with $D=1$, and c) homogeneous mixing ($D \rightarrow \infty$). 
While the prevalence $\rho$ is the same in the three cases, the number of active links (i.e. $S-I$ links) differs: one can see that infected particles are clustered together in the static case (a), while they become more homogenously distributed in the space as the diffusion constant increases (b) and particularly in the well-mixed case (c), where the number of active links reaches its maximum. 
In the following, we  address each case in detail.

\subsection{The static limit}
\label{ssec:static}

It is worth noting that, in the case of RnT walkers, the static limit can be obtained in two different ways, i. e,. $\alpha \to \infty$ at fixed $v$, or $v \to 0$ at fixed $\alpha$. For $D \to 0$, Eqs. (\ref{eq:model1}) and (\ref{eq:model2}) become $\partial_t S(\rr,t) = -\beta I(\rr,t) S(\rr,t) + \mu I(\rr,t)$ and $\partial_t I(\rr,t) = \beta I(\rr,t) S(\rr,t) - \mu I(\rr,t)$,
that have to be solved with the initial conditions $S_0=S(\rr,0)$ and $I_0=I(\rr,0)$. Because of the lack of  motion, the spreading dynamics can only occur in regions where the two populations have non zero overlap at $t=0$, i. e., $S_0 I_0 \neq 0$. The initial condition thus plays a fundamental role and, in order to study the properties of the stationary configurations, averaging over independent initial configurations is required.

In this case, the agent-based model is effectively described by a static network, where particles are represented by nodes, and two particles are connected by a link if they are within a distance $R$. 
Since particles do not move, links are fixed in time, i.e. the network is static.  
In particular, if initial conditions are random, i.e. particles are initially picked from a uniform distribution, the interaction network  is a random geometric graph \cite{dall2002random}.
The latter displays a percolation transition as the interaction radius $R$ is increased, with many microscopic connected components for low $R$ and the emergence of a macroscopic connected component at $R=R_c$. The interaction radius $R$ is related to the average connectivity $\langle k \rangle$ (average number of links per node) by
\begin{equation}
\langle k \rangle = (N-1) \frac{A_d(R)}{L^d}
\label{avdeg}
\end{equation} 
where $A_d(R)$ is the area of the $d$-dimensional sphere of radius $R$, being $A_2(R) = \pi R^2$, $A_3(R)=\frac{4}{3} \pi R^3$. 
The connectivity of the graph plays a key role on the emergence of an endemic phase: if the average degree of the graph $\langle k \rangle$ is above the percolation threshold $\langle k \rangle_c(d)$ (which crucially depends on the dimension $d$ of the system), then the graph displays a giant connected component that allows the emergence of macroscopic outbreaks.
For $\langle k \rangle \gg \langle k \rangle _c(d)$, the epidemic threshold will be the same as in the homogeneous-mixing regime,  $\beta_c/\mu = \langle k \rangle ^{-1}$ \cite{estrada2015random,rodriguez2019particle}.

The choice of the interaction radius $R$ is also relevant in the study of dynamical processes on motile agents as it defines the instantaneous underlying network structure. We set the interaction radius $R$ for all numerical simulations such that, in the  static limit, the average connectivity of the network is slightly above the percolation threshold, $\langle k \rangle \gtrsim \langle k \rangle_c(d)$.
 Hence, using eq. (\ref{avdeg}) and considering the critical connectivities ($\langle k \rangle_c=4.52$ for $d=2$ and $\langle k \rangle_c=2.74$ for $d=3$ \cite{dall2002random}), we fix  $R_2=0.04$ in $d=2$ and $R_3=0.089$ in $d=3$.



\subsection{The homogeneous-mixing limit}
\label{ssec:hom}

{The homogeneous-mixing limit} is recovered for $D\rightarrow \infty$, for a finite non-zero tumbling rate. In this limit, particles can travel an arbitrarily large distance in an arbitrarily short time interval. Their positions are thus effectively updated randomly. 
This means that each particle can, in principle, interact with any other within a small time interval, thus leading to the homogeneous mixing of the population.
From a network point of view, this case corresponds to an underlying contact network evolving much faster than the spreading process on its top, known as fast-switching or annealed network limit. 
In this limit, the underlying structure can be effectively approximated by a fully-connected graph, in which at each time step all particles may interact with every one else with a probability proportional to the average connectivity $\langle k \rangle$.
In this mean-field regime the SIS dynamics can be solved exactly \cite{RevModPhys.87.925}, and the epidemic threshold is  $\beta_c^{MF}/\mu = \langle k \rangle ^{-1}$.

\begin{figure}
\includegraphics[width=.52\textwidth]{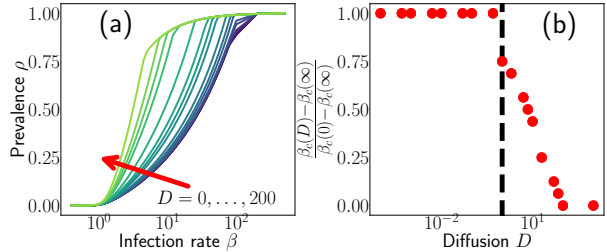}
\caption{Results for the continuum model in one dimension. (a) Prevalence $\rho$ as a function of the infection rate $\beta$, for different values of  $D$ (from $D\in[0,200]$, the red arrow indicates increasing values of $D$).
(b) Critical value $\beta_c$ as a function of $D$. For $D/\beta < 1$ (dashed black line) the transition is static-like, above this threshold, $\beta_c$ approaches the mean-field value, i. e., $\beta=\mu$, for increasing values of $D$.}
\label{figure:mod1d}
\end{figure}

\subsection{Static to mean-field limit crossover}
\label{ssec:cross}

As we stated before, from the point of view of the spreading dynamics, the system undergoes a crossover from a static limit, reached for $D\to0$,  to a mean-field regime obtained for increasing values of $D$. 
The key ingredient is thus the competition between the typical time scale of diffusion and that of the infection rate. The emerging phenomenology can be understood already at the level of the continuum model in one spatial dimension. 

To quantify this phenomenon, we consider the continuum model in the diffusive limit in a one-dimensional space, whose dynamics is governed by the following equations
\begin{align} \label{eq:model_1d}
    \partial_t S(x,t) &= D \partial_x^2 S + f(S,I) \\ 
    \partial_t I(x,t) &= D \partial_x^2 I - f(S,I)
\end{align}
with $x \in [0,L]$ and where we are assuming that the two species diffuse with the same diffusion constant $D$. We thus solved Eqs. (\ref{eq:model_1d}) numerically using Euler explicit scheme using periodic boundary conditions. We discretized the equations on a grid of $N_g=10^3$ points, with $\Delta x=1$ ($L=N_g$), and $\Delta t$ chosen in way such that $\Delta t \leq 1/2 D$. 
As control parameters, we move in $D\in[0,10^3]$ and $\beta\in[10^{-1},10^2]$. For the discrete Laplace operator, we adopted a standard finite difference method.
Moreover, we added to both initial condition $I(x,0)$ and $S(x,0)$ a small amount of noise and we averaged over $N_s=100$ independent noise realizations. 

The results from the numerical solution of the continuum model are shown in Fig. \ref{figure:mod1d} (here the prevalence is $\rho=\int \text{d}x \, S(x,\infty)$). As one can see, the functional form of $\rho$ as a function of $\beta$ depends on $D$ in a non-trivial way. 
We obtain that larger values of $\rho$ are reached sooner for large $D$ values. We can thus define $\beta_c(D)$ as the value of $\beta$ such that $\rho>5 \times 10^{-2}$. Once we identify  the static limit $\beta_c(0)$, we obtain that $\beta_c(D)=\beta_c(0)$ for $D/\beta < 1$, while  $\beta_c(D)$ starts to decrease (shown in Fig. \ref{figure:mod1d} (b)) when increasing the diffusivity. 
For large $D$ values, $\beta_c(D)$ approaches the mean-field limit $\beta_c(\infty) = \mu$ (we checked that $\beta_c(D)$ approaches a plateau for $D>50$ that does not change up to $D=10^{3}$). Although we are working in a very simplified picture, the one-dimensional model reproduces (i) the crossover between static to mean-field picture, and (ii) a diffusivity-dependent epidemic threshold $\beta_c(D)$, as obtained from the agent-based model described in the following section.

\section{Agent-based model}

\begin{figure}
\includegraphics[width=0.9\columnwidth,keepaspectratio]{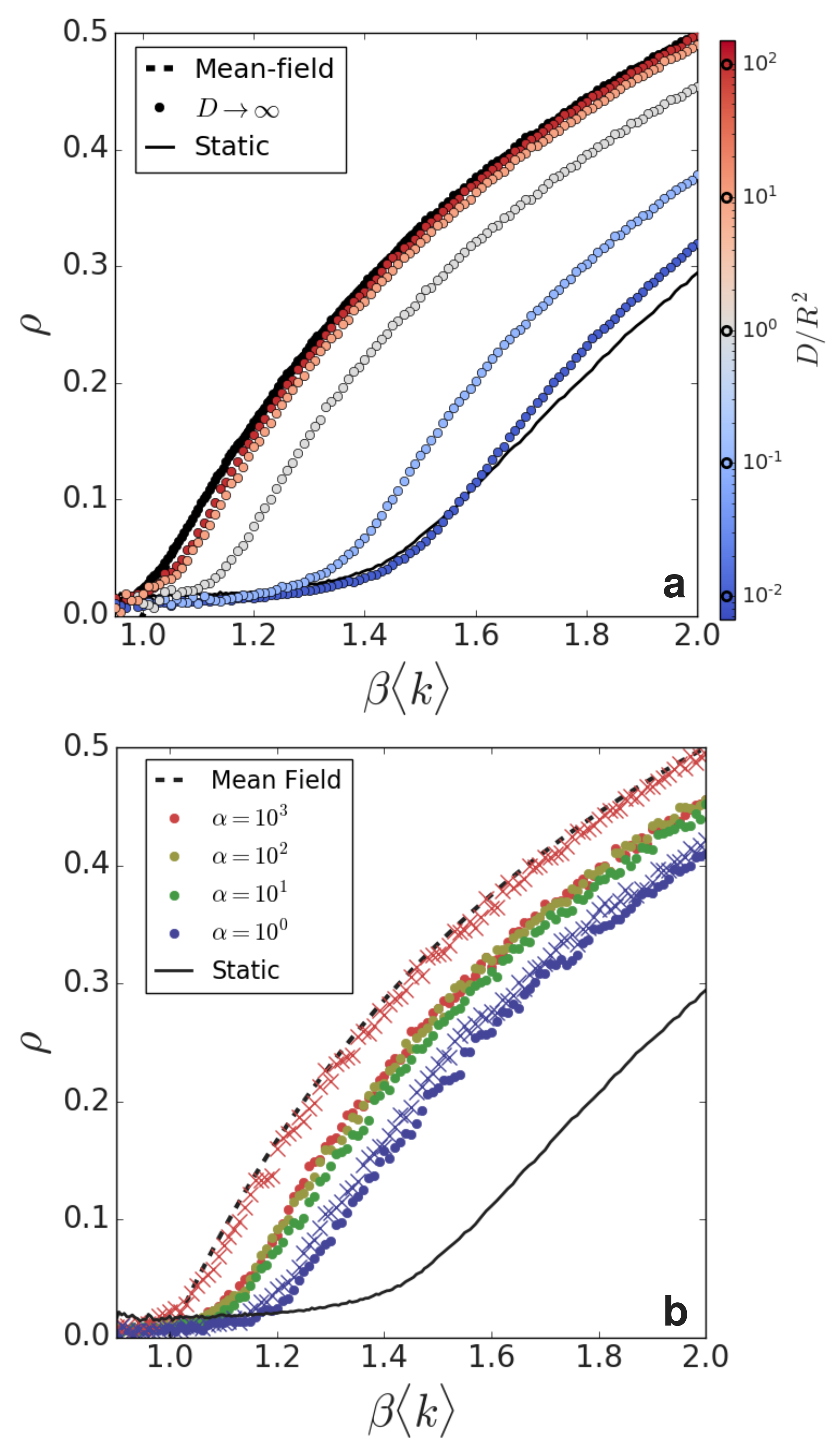}
\caption{ Prevalence $\rho$ versus the infection rate $\beta$, normalized with the inverse average degree  $\langle k \rangle$. Epidemic curves for: (a) different diffusion constants $D$, with fixed $\alpha=100$; (b) identical diffusion constant $D=\frac{v^2}{2\alpha}=R^2$ but different tumbling rates $\alpha$ (dots). Cross symbols correspond to purely ballistic agents moving at the velocity corresponding to the one of RnT with $\alpha=10^3$ (in red) and $\alpha=1$ (in blue). We also show for comparison the results obtained in the $D\rightarrow \infty$ (broken line) and $D=0$ (solid line) limits.}
\label{figure1} \label{fig2}
\end{figure}

In this Section we show results of numerical simulations of the agent-based model in $d=2$ and $d=3$, confirming the insights obtained from the continuum description.
Fig. \ref{fig2} shows the behavior of the agent-based model in  $d=2$.
The predictions of the continuum model are consistent with numerical simulations reported in Fig. \ref{fig2} (a), which shows the prevalence $\rho$ as a function of the rescaled infection rate $\beta \langle k \rangle$, for different values of the diffusion constant $D$ and a fixed tumbling rate $\alpha=100$. 
One can see that as $D$ increases, the epidemic curve approaches the mean-field (homogeneous-mixing) regime described in Section \ref{ssec:hom}, and illustrated by a dashed line in Fig. \ref{fig2} (a). 
In the same way, for $D \to 0$ the epidemic curves move towards the static limit (continuous line), described in Section \ref{ssec:static}.

In Section \ref{ssec:cross} we showed that, in the one-dimensional continuum model, the static to mean-field crossover is exclusively governed by the diffusion coefficient $D$ (Fig. \ref{figure:mod1d}). 
This picture is confirmed from the agent-based numerical simulations, see Fig. \ref{fig2} (b), which shows that the  curves $\rho(\beta,v,\alpha)$ with the same diffusivity $D=\frac{v^2}{2\alpha}$ but different values of $\alpha$ collapse to a single curve $\rho(\beta,D)$ (green, yellow, and red dots in Fig. \ref{fig2} (b)). 
Notably, this is true only in the diffusive regime, while for very small values of $\alpha$ (blue dots) the prevalence is different.  

Indeed, when $\beta$, the typical scale of the infection process, is higher than $\alpha$, infections occur more frequently than tumbles. This means that, on the time scale of the infection process, agents do not tumble but move only ballistically at velocity $v=\sqrt{2\alpha D}$. We confirm this picture analyzing the prevalence of purely ballistic agents, moving at the  velocity $v$ that a RnT agent would have for a given $\alpha\leq 1$, at fixed $D=R^2$ (Fig. \ref{fig2} (b)). Comparing the blue symbols in Fig. \ref{fig2} (b) for ballistic and RnT agents moving at same velocity, we confirm that for $\alpha\leq 1$, the SIS process spreads at the time scales of the ballistic component of RnT motion. On the contrary, comparing the red symbols in Fig. \ref{fig2} (b), ballistic and RnT agents at large $\alpha$ show a distinct behavior. The prevalence curve in the ballistic case is well different from the one obtained for RnT agents, being closer to the limiting homogeneous-mixing one. Thus, at large $\alpha$, the spreading process occurs at time scales where RnT motion becomes relevant and the crossover between static and mean-field behavior is controlled by the diffusivity.

%

\begin{figure}[tbp]
\includegraphics[width=\columnwidth,keepaspectratio]{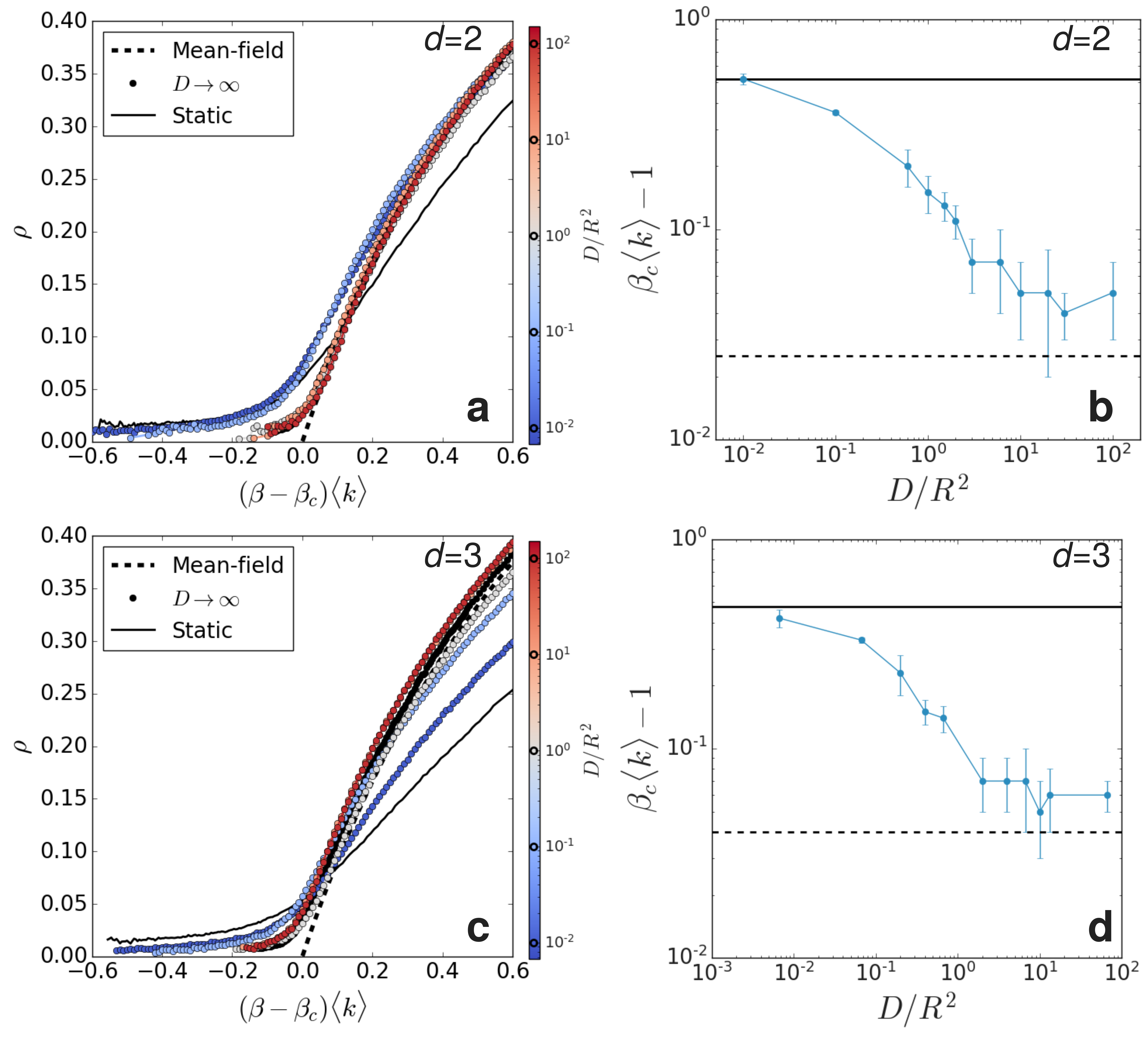}
\caption{Critical behavior of the agent-based model in two (panels a,b) and three (panels c,d) dimensions at fixed $\alpha=100$.
The prevalence $\rho$ as a function of the rescaled infection rate $\rho (\beta-\beta_c)\langle k \rangle$ in $d=2$ and $d=3$ is shown in panels (a) and  (c) respectively.
The difference between the epidemic threshold $\beta_c$   and the expected mean-field value $\beta_c^{MF} \langle k \rangle =\mu = 1$, as a function of $D/R^2$ is shown in panels (b) and (d).}
\label{f2d3d}
\end{figure}

Finally, we compare the critical behavior of the agent-based model in 2- and 3-dimensional simulation boxes of linear size $L$ with periodic boundary conditions.
We measured the epidemic threshold $\beta_c$ as the value that maximizes the susceptibility
\begin{equation}
    \chi(\beta\langle k\rangle)= N \frac{\langle \rho^2 \rangle - \langle \rho \rangle ^2}{ \langle \rho \rangle}
\end{equation}
where brackets in the right hand side denote averages over independent steady-states. 
Fig. \ref{f2d3d} (a), (c) show the prevalence $\rho$ as a function of the rescaled infection rate $\rho (\beta-\beta_c)\langle k \rangle$ in $d=2$ and $d=3$, respectively.
For most values of $D$,  the epidemic curves are very close to the behavior in the homogeneous-mixing, mean-field regime. Only for small values of $D\lesssim 1$ one can start seeing significant deviations from  mean-field, approaching the static behavior in the absence of motion as $D\to 0$. 
 
The value of  the ($D$-dependent) epidemic threshold $\beta_c$ is reported in Fig. \ref{f2d3d} (b), (d),  showing the difference between the epidemic threshold $\beta_c$ and the mean-field value, as a function of $D$.
As a reference, we also plot the actual threshold obtained from numerical simulations in the mean-field (dashed line) and the static  (continuous line) limit.
One can observe the crossover from static to mean-field described in Section \ref{ssec:cross}, the epidemic threshold $\beta_c$ decreases with $D$ and approaches the mean-field case continuously, in both two and three dimensions. 

Besides the dependence of the non-universal value of the epidemic threshold $\beta_c(D)$, diffusivity appears to control the crossover between different universality classes: the static one, corresponding to the SIS spreading on a finite dimensional network, and the homogeneous-mixing, mean-field one, corresponding to the SIS spreading in an infinite dimensional structure. Interestingly, beyond $D\approx R^2$, the emergence of a  finite fraction of infected agents is controlled by mean-field behavior and is thus largely independent of the underlying dimension of the space in which agents move.  Figure \ref{f2d3d} shows that for large enough $D$,  the curves $\rho$ vs. $\beta$ (once rescaled by $\beta_c$) are indistinguishable from the mean-field one within our numerical accuracy, giving support to the fact that the spreading mechanism in populations of fast enough diffusive agents is homogeneous, generically well captured by the homogeneous-mixing approximation.  

\section{Discussion and Conclusions}
In this work, we have studied the impact of motility on spreading dynamics. 
We established a general framework to tackle this problem, based on two paradigmatic models of both motility and spreading, namely,  Run-and-Tumble and SIS dynamics, bridging together active matter physics with epidemic spreading.  We restricted ourselves to the simplest case where interactions between agents a only mediated by the infection process, allowing to establish a field theoretic description. A natural extension of this problem would include pair-wise mechanical interactions, such as excluded volume, which are know to trigger clustering and patterning in systems of active particles \cite{Tailleur08,PhysRevX.7.011028, paoluzzi2020information}.
We focused in the diffusive limit, defined by a time-scale separation between  tumbling and spreading. 
This is a situation of practical interest if one is interested in taking into account the effect of mobility on large space and time scales. 

Within this framework, we obtained that the time-evolution of the density of infected and susceptible agents can be coarse-grained into a two-component reaction-diffusion equation conserving the total mass of the system. 
It is worth noting that, for small diffusion and large infection rate, the dynamics of the system is described by a set of coupled forward (for the susceptible) and backward (for the infected) diffusion equations with diffusion constants that are directly linked to the infection dynamics, i. e. $\tilde{D}_{S,I}=v^{2}_{S,I}/\beta$ in one spatial dimension. 
We notice that, although in general the backward diffusion equation is ill-defined and required some regularization, it has a simple physical meaning: it signals the tendency of the dynamics to make the profile of infected agents less smooth as time increases. This fact deserves future investigation. 

As a limiting situation, one has the so-called static limit, that is reached for vanishing values of the diffusion constant $D \to 0$. 
This static limit is quite intuitive: in absence of diffusion, regions with overlap of different species are the only ones where spreading can occur. 
This situation is well-represented by an epidemic process on random geometric graphs \cite{RevModPhys.87.925}, where the outbreaks will only occur in the connected components with at least one initially infected particle.
On the other hand, as the diffusion becomes important, the model undergoes a crossover towards mean-field behavior, and thus the system reaches the so-called homogeneous-mixing limit, in which all particles can, on average, interact with all the others.
We tested and documented the presence of this crossover by solving numerically the continuum model in one spatial dimension. 
The mean-field regime is obtained because the initial density profiles of infected and susceptible agents relax towards the homogeneous profile on a typical time scale  smaller than the one associated to the SIS process. 
We observed that the critical value $\beta_c$ is diffusion-dependent, i. e., $\beta_c=\beta_c(D)$, and continuously switches between two regimes: (i) for small $D$ values, $\beta_c$ is $D$-independent, basically controlled by the behavior of the SIS model on a static short-ranged network, while (ii) it decreases to smaller $\beta_c$ as $D$ is increased. 

We tested the predictions of the continuum model against numerical simulations of SIS epidemic process on top of run-and-tumble walkers in two and three dimensions, showing the presence of the crossover from a mean-field regime, that is reached at high diffusion constant, to the static limit for small diffusion values. 
The nature of such crossover is revealed by the behavior of the local prevalence that is characterized by localized spots at zero diffusion that become more and more extended for increasing value of the diffusion constant.  Above a threshold $D_c\approx R^2$, the transition towards a state with non-zero prevalence appears to be of the mean-field kind. While the specific location of the epidemic threshold does depend on $D$, at large enough diffusivities, the transition is controlled by the mean-field, homogeneous-mixing, regime.

As a future direction, it might be interesting to explore different motility regimes, e. g., when the persistence time is of the same order as the SIS time scales, allowing to investigate the role of persistent motion on the spreading dynamics \cite{paoluzzi2021single}.
It might be also interesting to study pattern formation driven by feedback between the SIS state variable and motility parameters \cite{PhysRevE.98.052603,paoluzzi2020information}.

\section*{Acknowledgments}
M.P. has received funding from the European Union's Horizon 2020 research and innovation programme under the MSCA grant agreement No 801370 and by the Secretary of Universities 
and Research of the Government of Catalonia through Beatriu de Pin\'os program Grant No. BP 00088 (2018). J.P.R. is supported by Juan de la Cierva Formacion program (Ref. FJC2019-040622-I) funded by MCIN/AEI/ 10.13039/501100011033. D.L. acknowledges MCIU/AEI/FEDER for financial support under Grant Agreement No. RTI2018-099032-J-I00.
M.S. acknowledges support from Intesa Sanpaolo Innovation Center.
\bibliography{mpbib}

\appendix 

\end{document}